# Neural geometry in the human hippocampus enables generalization across spatial position and gaze

Assia Chericoni[1], Chad Diao[1], Xinyuan Yan[1], Taha Ismail[1], Elizabeth A. Mickiewicz[1], Melissa Franch[1], Ana G. Chavez[1], Danika Paulo[1], Eleonora Bartoli[1], Nicole R. Provenza[1,2], Seng Bum Michael Yoo[3], Jay Hennig[2,4], Joshua Jacobs[5], Benjamin Y. Hayden[1,2]*†, Sameer A. Sheth[1,2]*

1. Department of Neurosurgery, Baylor College of Medicine, Houston, Texas, United States of America
2. Department of Electrical and Computer Engineering and Neuroengineering Initiative, Rice University, Houston, Texas, United States of America
3. Department of Biomedical Engineering and Department of Intelligent Precision Healthcare Convergence, Sungkyunkwan University, Suwon, Rep. of Korea
4. Department of Neuroscience, Baylor College of Medicine, Houston, Texas, United States of America
5. Department of Neurology, University of Chicago, Chicago, Illinois, United States of America

* These two authors contributed equally
† Correspondence: Benjamin.Hayden@bcm.edu

**Funding statement**
This research was supported by the McNair Foundation, by NIH R01 MH129439, and NIH R01 DA038615.

**Competing interests**
S.A.S has consulting agreements with Boston Scientific, Zimmer Biomet, Koh Young, Abbott, and Neuropace. SAS is Co-founder of Motif Neurotech

**Acknowledgements**
We thank Joshua Adkinson, Justin Fine, Victoria Gates, George Kokalas, Raissa Mathura, and Andrew Watrous for invaluable assistance.

# ABSTRACT

Hippocampal neurons track positions of self, others, and gaze direction. However, it is unclear how their respective neural codes differ enough to avoid confusion while allowing for abstraction. We recorded from populations of hippocampal neurons while participants performed a joystick-controlled virtual prey-pursuit task involving multiple moving agents. We found that neurons have mixed selective responses that map positions of self, prey, and predator, as well as gaze. Their codes occupied mostly orthogonal subspaces, but these subspaces' geometric structure allowed them to be aligned by simple linear transformations. Moreover, their geometry supported generalization across spatial maps, such that a linear rule learned on one agent transfers to another. This scheme enables reliable individuation and abstraction across both agent identity and viewpoint. Together, these findings suggest that hippocampal spatial knowledge is structured as a family of geometrically related manifolds that can be flexibly aligned to different agents and gaze directions.

**INTRODUCTION**

As we move around the world, we need to keep track of where we are and where other important individuals are in the same environment (Ekstrom et al., 2018; Epstein et al., 2017; Moser et al., 2017). This tracking can be implemented with the help of neural maps, that is, firing rate-driven representations of position (O'Keefe & Dostrovsky, 1971; Moser et al., 2008). For example, hippocampal place cells, which exhibit increased firing rates when the individual enters a specific spatial location, play a key role in tracking self-position. Likewise, social place cells track the positions of key other individuals (Omer et al., 2018; Danjo et al., 2018). However, current theories do not readily account for how we simultaneously track self and other, or two different others. At first blush, the problem appears trivial: we can solve it with separate neurons (e.g., labelled line codes) or orthogonal coding axes. However, this solution does not account for our ability to generalize across agents and contexts. Consider, for example, a child playing tag. While running, she sees a friend slip on a muddy spot on the floor. Even without stepping there herself, she infers that the floor is slippery and adjusts her trajectory to avoid it. This capacity for cross-individual generalization is a hallmark of human cognition, yet it sits in tension with the need to distinguish representations of self and others (Behrens et al., 2018; Park et al., 2020; Thornton and Mitchell, 2018).

Along with its role in spatial mapping, the hippocampus has also been implicated in representing where we look (Ekstrom et al., 2003; Jutras et al., 2013; Killian et al., 2012; Hoffman et al., 2013; Leonard et al., 2015; Leonard & Hoffman, 2017; Meister & Buffalo, 2018). However, there is ongoing debate about the nature of hippocampal gaze tracking. Some accounts emphasize that hippocampal responses primarily reflect the observer's physical position in the world, whereas others argue that they primarily encode gaze location or attentional sampling of space (Mao et al., 2021; Martinez-Trujillo, 2025; Nau et al., 2018; Payne & Aronov, 2025; Piza et al., 2024; Watrous & Ekstrom, 2014; Wirth et al., 2017; Yoo et al., 2020). Resolving this issue is challenging because gaze and position are typically tightly coupled during natural behavior (Epstein et al., 2017). Importantly, if the hippocampus represents both self-location and gaze location, it faces a problem analogous to that posed by multi-agent tracking, namely, how to maintain separable representations while preserving shared structure across reference frames. This generalization problem is quite serious - consider that we must recall that the world stays the same regardless of our vantage point (Melcher and Colby, 2008). How the hippocampus might simultaneously encode body-centered and gaze-centered information, without interference yet with the capacity for generalization, remains unknown.

One potential solution to these representational challenges comes from recent advances in theories of neural manifold coding (Ebitz and Hayden, 2021; Langdon et al., 2023; Perich et al., 2025; Vyas et al., 2020). Rather than assigning each variable to independent neurons or orthogonal coding axes, population activity can be placed on a structured, low-dimensional manifold that can be reused across related conditions but is separable when necessary (Barak and Fusi, 2013; Bernardi et al., 2020; Johnston et al., 2024). Such codes have proven useful in explaining flexible decision-making, motor control, working memory, and context-dependent behavior (Alleman, 2024; Cunningham & Yu, 2014; Elsayed et al., 2016; Fine et al., 2023; Gallego et al., 2017; Libby and Buschman, 2021; Parthasarathy et al., 2017; Tang et al., 2020; Xie et al., 2022; Yoo and Hayden, 2020). Indeed, a small but growing body of evidence indicates that population geometry principles apply to hippocampal spatial codes (Esparza et al., 2023; Fenton et al., 2010; Fenton, 2024; Levy et al. 2023; Low et al., 2018; Nieh et al., 2021; Tang et al., 2023). According to this view, locations of different individuals or gaze can be encoded as partially overlapping maps, implemented in orthogonalized or semi-orthogonalized linearly transformable subspaces within a common representational geometry. Such an organization could, in principle, allow the brain the ability to generalize across while still keeping these signals distinguishable (Bernardi et al., 2020). We hypothesized that the hippocampus represents self-location, others-location, and gaze using semi-overlapping population maps arranged as semi-orthogonal linearly transformable subspaces, providing a unified framework that supports both generalization and differentiation across individuals and perspectives.

To test how the hippocampus resolves these representational demands, we examined single-neuron activity in humans performing a continuous, naturalistic navigation task in a virtual open field (Yoo et al., 2020). Human participants used a joystick to pursue moving prey while, on some trials, simultaneously avoiding a predator, creating a setting in which the positions of self, multiple others, and gaze were all behaviorally relevant and dynamically changing. We found that individual hippocampal neurons carried information about the positions of the self-avatar, prey, predator, and gaze, and used mostly (but not completely) distinct response fields for each. At the population level, these signals were organized into semi-orthogonalized subspaces linked by a consistent linear relationship. This linear relationship preserves internal geometry, thereby embodying a structure that can support *abstraction* (using the term in the same sense as Bernardi et al., 2020). We therefore functionally tested for abstraction using cross-condition generalization performance (CCGP). We found that linear decoders trained to read out spatial variables for one agent or viewpoint successfully transferred to others, indicating a *generalizable* representation (Bernardi et al., 2020). Together, these results suggest a "both" solution to the longstanding debate about whether primate hippocampus specializes in gaze or position (Martinez-Trujillo, 2025). More broadly, they support a view of the hippocampus as implementing a family of semi-overlapping spatial manifolds that can be flexibly re-centered on different agents and viewpoints.

**RESULTS**

**Prey pursuit in humans**

　　Human participants (n=21) performed the *prey-pursuit task* (**Figure 1A**, **Methods;** Chericoni et al., 2025). On each trial, the participant used a joystick to continuously move the position of an avatar (yellow circle) in a rectangular field displayed on a computer screen (**Supplementary Video, Figure 1A** and **B**). The participant had up to 20 seconds to capture fleeing prey (colored squares) to obtain points. Prey avoided the avatar with a deterministic strategy that combined repulsion from the avatar's current position with repulsion from the walls of the field (**Methods**). The prey items were drawn randomly on each trial from a set of three that differed in maximum velocity and reward size. Over the course of each session, participants' avatars typically spanned the entire field (**Figure 1B**).

　　Each trial began with one or two prey appearing at one or two of the cardinal points (**Figure 1A**). In two-prey trials (from 50% to 81% of trials, depending on the participant), the participant was free to decide which prey to pursue at any moment (**Figure 1C, D**). In trials with two prey, participants successfully captured one of the prey in 71.25% of trials (**Figure 1E**), and, on successful trials, did so in an average of 7.51 seconds (variance: 1.89 seconds; **Figure 1F**) with an average reaction time of 0.87 seconds (standard error: 0.07 seconds), defined as the interval between the agents' appearance on the screen and the participant's first move. Here and throughout, we use *agent* simply to refer to the self avatar and the prey, as a neutral descriptive label. Participants' performance did not depend significantly on prey type (trial length x reward level, p = 0.07; reaction time x reward level, p = 0.24). In five participants, we employed a variant of the task in which, in addition to the prey, there was also a predator pursuing the participant's avatar (**Figure 1D**). In this case, the predator used a simple distance-minimizing pursuit strategy (**Methods**). Our patients successfully evaded capture by the predator in 95.2% of trials.

　　We recorded responses of 726 neurons in the hippocampus while participants performed this task (average n=34.6 neurons per participant, **Figure 1G**). Of these neurons, 176 were recorded in the variant of the task with a predator. **Figure 1H** shows responses of two neurons with clear event-aligned responses; overall, however, neurons had complex selectivities that were not easily described as being about trial start and stop. We therefore developed a more sophisticated approach to understanding how neural responses related to the positions of avatars in the task.

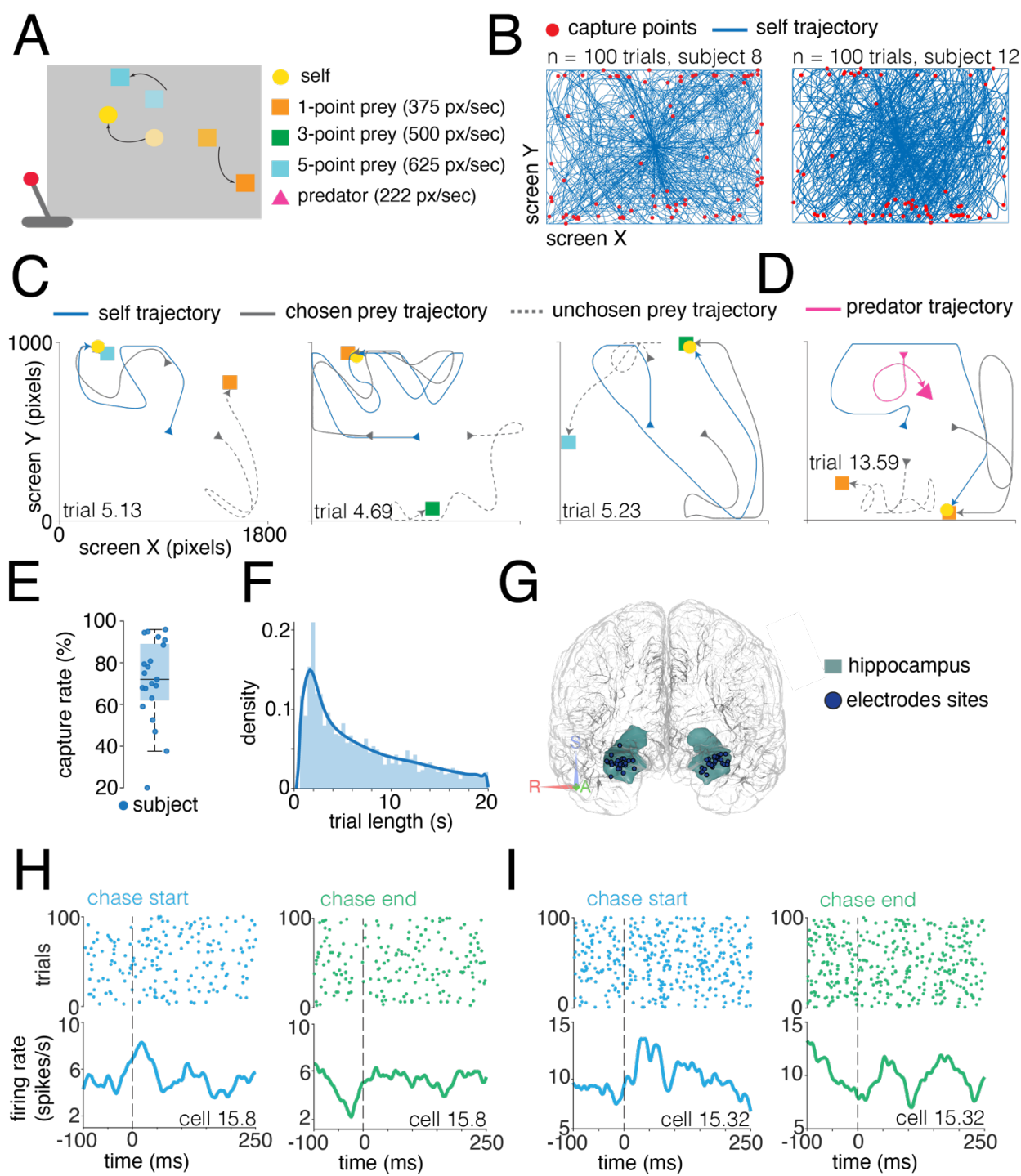

**Figure 1. Task design and neural recordings. A.** Schematic of pursuit task design. Prey were displayed as colored squares that denote their reward if caught and maximum speed. Participants' avatar was displayed as a yellow circle, with avatar velocity determined by the magnitude of the joystick deflection from its center. For some participants, we also included a predator (**Methods**). **B.** Example sessions from two participants; blue lines indicate participants' trajectories on each trial overlaid; red dots indicate points of prey capture. **C and D.** Typical example trials; arrowheads indicate direction of travel. Trials are identified as *participant*

*number.trial number.* Blue line: self-avatar trajectory; gray line: chosen prey trajectory; dashed gray line: unchosen prey trajectory; magenta line: predator trajectory. **E.** Distribution of performance (proportion of trials in which prey was captured) across our 21 participants.**F.** Histogram of trial durations; each trial ended either with the successful capture of the prey or after 20 sec. **G.** Glass brain showing recording location of neurons in the human brain for all patients. **H and I.** Example single neuron responses from two neurons aligned to trial start (blue) and trial end (green). Dashed vertical lines indicate the beginning and the end of the trial. Plots display mean firing rates.

**Hippocampal maps for positions of self, prey, and predators**

     To estimate spatial mapping functions in these neurons, we adopted the Poisson generalized linear model (GLM) procedure developed by Hardcastle et al. (2017). This approach fits response profiles with minimal *a priori* assumptions about the shape of the tuning surface (**Methods**). Using this approach, we found 29.5% (n=214/726) of neurons track the position of the self-avatar on the monitor (**Figure 2A**). Because the participants freely chose one prey to pursue, the two different prey could be differentiated based on their status as either *chosen* or *unchosen* (defined as whether they were eventually captured, **Figure 1C, D**). We found that 26.3% of neurons (n=191/726) track the position of the chosen prey, and 19.0% (n=138/726) track the position of the unchosen prey (**Figure 2A**; both proportions are greater than chance, binomial test, p<0.05). In the neurons that we recorded during predator trials, 33.0% (n=58/176) track the position of the predator (**Figure 2A**).

     We found no evidence for discrete populations of neurons corresponding to the different agents. Indeed, these populations were broadly overlapping, indicating that, if a neuron participated in mapping one agent, it was likely to participate in mapping multiple agents (**Figure 2B**). Specifically, 12.0% (n=88/726) of the neurons encode the position of any two agents, and 10.5% (n=76/726) map all three agents. Similarly, during predator trials, 29.0% of the neurons (n=51/176) track the position of two or three agents, with 10.2% (18/176) track all four agents. We evaluated the quality of fit of each neuron's firing structure through correlation (Pearson's) between the observed firing rates and those predicted by the GLM (**Figure 2C**). Correlations were positive (mean = 0.12), confirming that the estimated tuning profiles capture meaningful spatial structure.

     Example tuning surfaces are shown in **Figure 2D**. These surfaces illustrate the range of response fields we observe (additional maps are shown, in heatmap view, in **Figure 2E**). The observed distribution of tuning peaks was generally broad, and any central hotspot was often accompanied by other, often lower amplitude, peaks. Together, these results indicate that, while a minority of neurons showed activity that resembles classic place cell activity in its shape, they are part of a larger and more heterogeneous distribution of tuning functions (cf. Hardcastle et al., 2017; Harland et al., 2021; Varga et al., 2024).

     Next, we asked whether hippocampal neurons have different maps for the different agents (**Figure 2F and G**). To quantify the relationship between maps, we used a *spatial similarity index* (SPAtial EFfieciency metric - SPAEF, Koch et al., 2018; **Methods**), which

measures the correlation between spatial representations. A value of zero indicates full orthogonality between maps, while values closer to +1 or -1, respectively, indicate correlation or anti-correlation between maps. The mean spatial similarity between self and chosen prey maps is 0.13. This value is greater than zero ($p < 0.001$, **Methods**) and below noise ceiling (0.22, quantified by half-splits bootstrapping), meaning it is significantly orthogonal ($p < 0.001$). Together, these results indicate that these maps are not randomly related; instead, they are weakly positively correlated. Indeed, the mean spatial similarity between self and unchosen prey maps and between chosen and unchosen prey maps is also greater than zero, again by a small amount (self - unchosen prey SPAEF = 0.04, $p < 0.001$; chosen - unchosen prey SPAEF = 0.006; $p < 0.001$). Not surprisingly, all SPAEF values were below noise ceiling (self - unchosen prey ceiling = 0.21; chosen - unchosen prey ceiling, SPAEF = 0.19; $p < 0.001$). In the subset of neurons in which we had predator data, we found that the spatial similarity between self and predator was slightly negative (SPAEF =-0.01, $p < 0.001$).

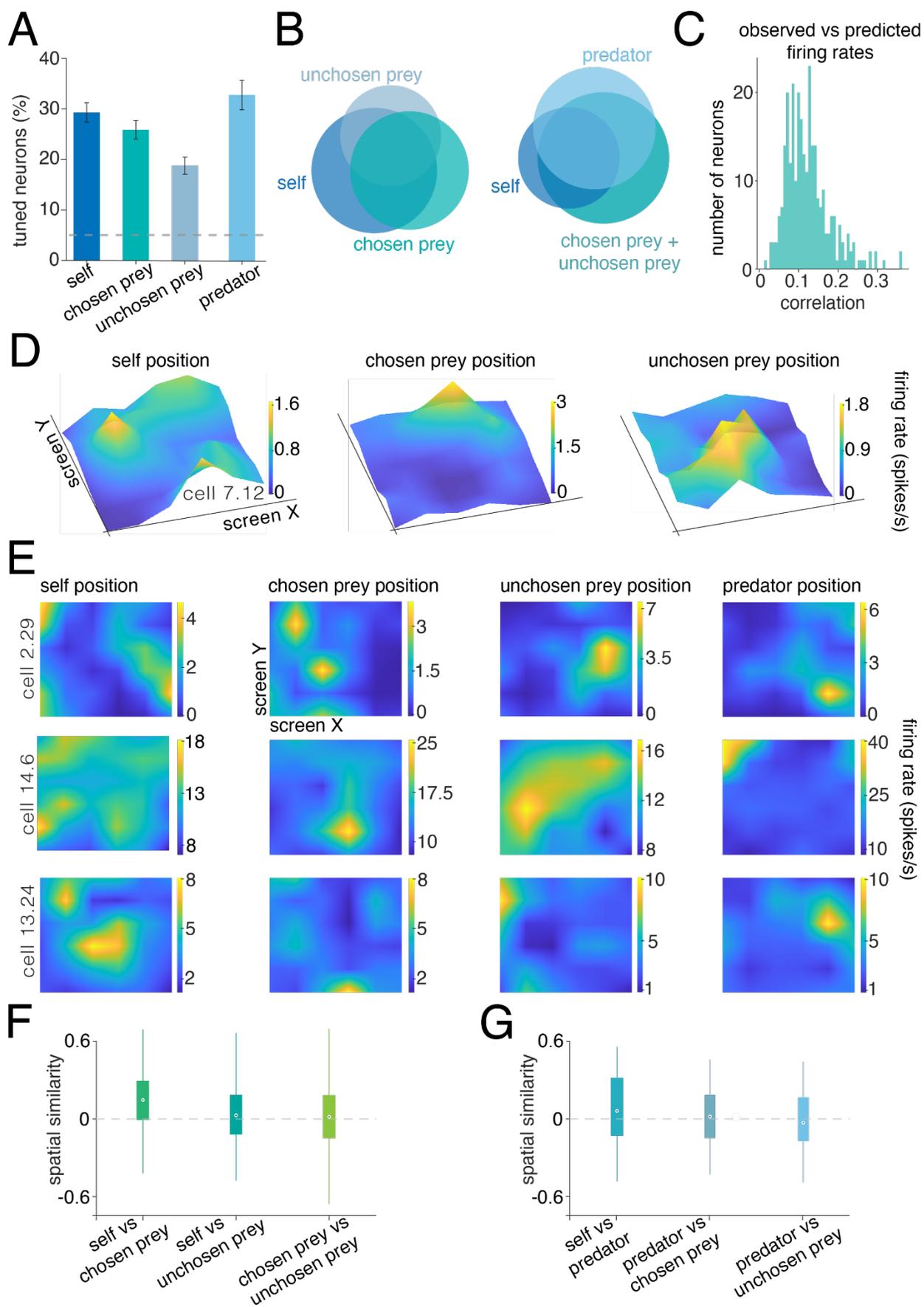

**Figure 2. Hippocampal mapping functions for self and prey. A.** Bar plot showing the proportion of neurons with significant tuning for agent positions according to the GLM. Proportions do not sum up to one hundred as one neuron may be tuned to more than one agent's position. Dashed horizontal gray line: 5% chance level. Error bars: standard errors. **B.** Venn diagram showing overlap between neurons selectivities. Neurons generally encoded positions of multiple agents. **C.** Distribution of correlations between observed and GLM predicted firing rates for significantly tuned neurons only. Positive correlations indicate that the GLM reliably captured each neuron's spatial firing pattern. **D.** Heatmaps showing tuning of three example neurons to the positions of the self, chosen prey, and unchosen prey. Neurons often showed a single hotspot but typically showed more complex features. **E.** More examples of tuned neurons of each type. Each row corresponds to a single neuron, while each column represents the firing activity relative to different agents (self, chosen prey, unchosen prey, and predator when present). Yellower regions indicate locations where the neuron exhibited higher firing rates. **F and G.** Boxplots representing the median spatial similarity between maps across different agent representations. Error bars: standard error. Spatial similarity values near zero indicate orthogonality (low similarity) between maps. All the neurons are shown.

**Hippocampal spatial maps occupy separable yet linearly related subspaces**

To assess the degree of orthogonality for subspaces for each individual, we modified a method previously developed to study motor regions (Elsayed et al., 2016). This analysis quantifies the degree of overlap between population representations, here applied to spatial selectivity for self and others, by projecting the population activity onto low-dimensional subspaces that capture most of the variance between agents' representations (**Methods**).

We first computed the correlation matrices for self, prey and predator tuning functions (**Figure 3A**). These matrices are, essentially, a plot of all possible correlation coefficients between the tuning functions for all possible pairs of neurons. In the first row of **Figure 3A** the entries of the matrix are sorted by dendrogram with centroid linkage, creating a hotspot in the center. The critical finding here is that when sorting the other matrices using the same indices, this hotspot nearly disappears, indicating that correlation *across* matrices is close to zero. In other words, maps for self and others are almost entirely orthogonal. Indeed, when we compared the entries of the self matrix against the entries of the prey and predator matrices, as well as the entries of the chosen prey matrix against the ones of the unchosen prey matrix, we found little correlation (self - chosen prey $R^2 = 0.02$; self - unchosen prey $R^2 = 0.008$; self - predator $R^2 = 0.01$; chosen - unchosen prey $R^2 = 0.007$; all p < 0.001; **Figure 3B**). While these correlations reached statistical significance, the effect size is small, indicating that the spatial representations were largely distinct.

Why are these tuning functions (mostly) uncorrelated? One possibility is that these neurons form distinct pools, each selectively representing one agent only (this would be called a *labelled line* code). However, given the overlapping maps observed in our tuning analysis (especially **Figure 2B**), we expect this not to be the case. To confirm this prediction, for each neuron, we calculated an agent-preference index (API) that quantifies the modulation of the neural activity for one agent compared to the modulation for another agent (this is what Elsayed et al. call the epoch preference index). A neuron that is concerned with the representation of one or another agent would have an index of +1 or -1, while a neuron selective for multiple agents would have an intermediate index of 0. Therefore, the presence of self-, prey-, and predators-only neurons would result in a bimodal distribution of APIs across the population of neurons. All the APIs distributions (for each couple of agents) cluster tightly around zero. Indeed, we can statistically test for bimodality (Hartigan dip test for bimodal distribution, all p = 0.99; histograms in **Figure 3B**). As a result, any differences between these representations are likely to come from how their activity patterns are arranged relative to each other in population subspaces.

To identify the relationship between these activity patterns, we used a principal component analysis (PCA)-based approach (Elsayed et al., 2016; Johnston et al., 2024). If the coding subspaces for self and others are orthogonal, the self-PCs should capture little to no prey and predator variance. Indeed, we found that the top ten self-PCs explained 70.7% of the variance of self maps, but they captured very little prey and predator maps variance (12.4% of chosen prey variance; 9.6% of unchosen prey variance; 12.6% predator variance; **Figure 3C, D**). We validated these results with half-split cross-validation (CV, **Methods**). For each split, we randomly subsampled half of the self- maps and computed a self covariance matrix. We then computed an analogous covariance matrix for matched subsampled maps for prey and predators and correlated these covariance matrices (Pearson's correlation), repeating the process 500 times. This yielded a distribution of within-self pairwise correlations (self-self noise floor) and corresponding distributions of across-agent (self-prey, self-predator) correlations on matched splits. Across splits, self-prey correlations did not exceed within-self noise floor (self-chosen prey $R^2 = 0.008$, p = 0.48; self-unchosen prey $R^2 = 0.005$, p=1). The chosen-unchosen prey and self-predator correlations were small, but significantly higher than noise baseline (chosen-unchosen prey $R^2 = 0.005$, p < 0.001; self-predator $R^2 = 0.02$, p = 0.02).

We next computed the *alignment index*, which measures the degree of rotation between two subspaces. Alignment index is defined as the variance of prey and predator tuning functions captured by the top ten self-PCs, normalized by the variance captured by the top ten PCs for each corresponding agent (**Methods**; Elsayed et al., 2013; Yoo et al., 2020). Values near zero indicate orthogonal subspaces, whereas values near one indicate strong alignment. We found that the alignment between self and the other agents was greater than zero (i.e. at least somewhat collinear) and significantly orthogonal (self-chosen prey = 0.15; self-unchosen prey = 0.12; self-predator = 0.17; all p < 0.001; **Figure 3E**). The chosen-unchosen prey alignment was also low (0.13; p < 0.001). Thus, self and others representations lie in semi-orthogonal (meaning, between orthogonal and collinear) neural subspaces.

**Subspaces are distinct but linearly transformable**

This semi-orthogonal structure suggests that, despite the mixed selectivity at the single neuron level, the spatial representations may be organized in a way that allows self and other maps to be separated at the population level. That is, spatial maps for different agents are not merely entangled but rather occupy distinct directions in the neural space, which makes their population variance separable. If this is true, we should be able to isolate dimensions that capture variance specific to each map.

Importantly, such separation would not be achievable if the spatial maps were fully collinear. Therefore, we solved a generalized eigenvalue problem that identifies the dimensions capturing self variance and then constructs an orthogonal set of dimensions for prey and predator (Boumal et al., 2014; Cunningham & Ghahramani, 2014; Elsayed et al., 2016; Yoo et al., 2020; **Methods**). We found that each agent's spatial representation contains self-, prey- and predator-specific dimensions that can be arranged into mutually orthogonal subspaces (**Figure 3F**). Specifically, we identified four self- dimensions that captured 38.1% of the self maps variance; four dimensions for the chosen and unchosen prey, that capture respectively 38.7% and 41.6% of chosen and unchosen prey maps variance; and four predator dimensions that explained 44.9% of the predator maps variance (**Figure 3F**). (We used four dimensions per agent, as this provided a balance between variance explained and cross-validated decoding performance, **Methods**). Each self, prey, and predator subspace was specific to its corresponding agent: projecting self maps onto the prey and predator subspaces (and vice versa) explained only a minimal portion of variance (bootstrap, all $p > 0.05$; **Figure 3F**). This analysis shows that spatial representations for self, prey and predators, that are mixed at the single neuron level, are separable at the geometry level.

Orthogonal subspaces are not necessarily totally unrelated. Two subspaces can be orthogonal but can, if structured appropriately, have a systematic relationship defined by a linear transformation (such as a rotation). This linear transformability allows the subspaces to be readily translated when generalization is favored but kept distinct when it is not (Kaufman et al., 2014; Elsayed et al., 2016; Yoo and Hayden, 2020). To test for linear transformability between subspaces, we trained a linear decoder to predict one agent's spatial map from another's within the corresponding orthogonal subspaces (**Figure 3G-I**). Specifically, we used self maps projected onto the 4-dimensional self-subspace to predict the activity patterns in the chosen and unchosen prey subspaces (**Methods**). Self and prey maps were indeed related through a linear mapping (self-chosen prey $R^2 = 0.40$, **Figure 3G**; self-unchosen prey $R^2 = 0.38$). Leave-one-out cross-validation (LOOCV) confirmed that predictions on held out conditions remained significantly above the shuffled baseline (self-chosen prey LOOCV $R^2 = 0.29$, $p < 0.001$, **Figure 3G**, see insets; self-unchosen prey LOOCV $R^2 = 0.24$, $p < 0.001$, data not shown). Predator maps were also linearly predictable from self maps ($R^2 = 0.46$; LOOCV $R^2 = 0.28$, $p < 0.001$; **Figure 3H**). Finally, chosen and unchosen prey subspaces were strongly linearly related ($R^2 = 0.92$; LOOCV $R^2 = 0.85$, $p < 0.001$; **Figure 3I**). These results indicate that while the hippocampus uses separate codes for different agents, these codes are linearly transformable.

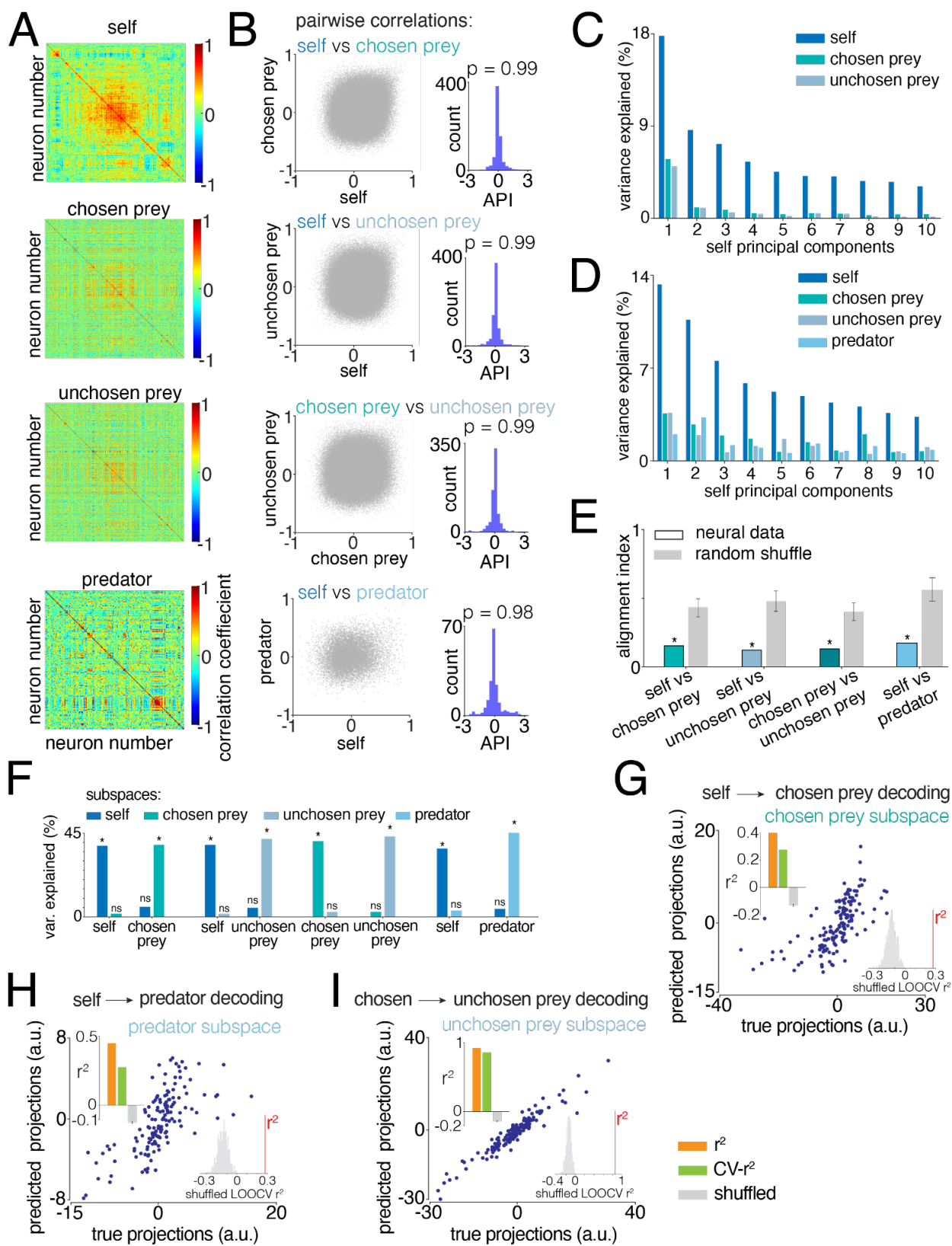

**Figure 3. Population level organization separates and links spatial maps for self and others. A.** Correlation matrices for self, prey, and predator tuning functions across neurons.

Each X and Y axis represents individual neurons, and color (Z-axes) indicates the Pearson correlation coefficient between pairs of neurons. The order of the neurons is identical across matrices, determined by hierarchical clustering of the self matrix using centroid linkage. The correlation structure is different across matrices. **B.** Relationship between pairwise correlation values across each pair of matrices (on the left). Each point represents a neuron pair. The weak correlation indicates that the similarity structure among neurons for self does not predict similarity structure among neurons for prey and predators. The histogram (on the right) shows the distribution of the *agent preference index*, quantifying the strength of neural activity for self compared to that for others. All the distributions are not significantly bimodal (*p* indicates Hartigan's dip test, all p > 0.05). **C.** Percent of self, chosen, and unchosen prey variance explained by the top ten self-PCs. **D.** Percent of self, prey, and predator variance explained by the top ten self-PCs. **E.** Alignment index for neural and random data. Self and others representations are more orthogonal than expected by chance. **F.** Percent of variance explained by the self and prey subspaces. Cross-maps projections significantly reduce the variance explained. **G.** Chosen prey subspace projections predicted from self subspace coordinates using a linear transformation, plotted against the observed chosen prey projections. Each point represents a spatial condition (36 spatial bins across the 4-dimensional subspace). The histogram in the inset shows the shuffled cross-validated $R^2$ distribution; significance demonstrates that self and chosen prey subspaces are related by a linear transformation. Decoding performance is significantly greater than chance. The bar plot in the inset shows the linear decoder performance: training $R^2$, leave-one-out cross-validation $R^2$, and shuffled control for self-chosen prey linear mapping. Error bars denote 95% confidence interval of the shuffled distribution. **H.** Same as in G, but for predator projections predicted from self subspace coordinates. **I.** Same as in G, but for unchosen prey projections predicted from chosen prey subspace coordinates.

**Distinct yet complementary coding of gaze and self position**

To assess gaze direction selectivity, as well as its relationship with other spatial codes, whether hippocampal activity reflects gaze direction in addition to spatial position, we recorded eye movements from six patients while they were performing the prey pursuit task (**Figure 4A-B** and **Methods**). We used the same GLM and subspace frameworks applied previously to self, prey and predator representations. As with the different avatars, gaze direction is correlated with the positions of the prey during task performance, so we leveraged the ability of the GLM to deal with the resulting multicollinearity (Hardcastle et al., 2017).

First, we asked whether individual hippocampal neurons showed responses that were modulated by gaze position, either independently or together with other spatial variables. Using self, gaze, prey positions and their combinations as predictors, we found that a comparable proportion of neurons encoded gaze position (17.1%, n = 26/152; **Figure 4C**) and self position (18.4%, n = 28/152; **Figure 4C**). We next focused on the subset of neurons collected from patients with sufficiently large number of trials for follow-up analyses (n=4, see **Methods**). In this subset of participants, a significant fraction of neurons was also tuned to the position of the chosen prey (12.5%, n = 19/152; **Figure 4C**) and unchosen prey (8.6%, n = 13/152; **Figure 4C**). As above, we found that these populations were not segregated and showed some degree of overlap: 10.3% of neurons (n = 16/152; **Figure 4D**) encoded the combination of two or three predictors.

We next asked whether gaze firing rate modulation merely recapitulates spatial tuning for self position (Martinez-Trujillo, 2025). If apparent gaze coding were an artifact of the correlation between gaze direction and self location, then including gaze as a predictor in the GLM should improve the encoding of self position. Conversely, if gaze carries an independent signal, adding it to the model should have little impact on self tuning. We therefore refit the GLM with and without gaze as a predictor. Including gaze did not substantially improve model performance for self or prey tuning, indicating that gaze position is not a trivial reflection of body position but instead contributes distinct information to hippocampal firing (**Figure 4E**). Model fit quality, measured as the Pearson correlation between observed and GLM predicted firing rates, was comparable across models (with gaze: r = 0.11; without gaze: r = 0.11). For single neurons, gaze tuning was spatially structured and comparable to self, prey, and predator tuning (**Figure 4F**, unchosen prey and predators data not shown).

We next asked how self and gaze maps are organized relative to each other at the population level. To visualize their structure, we computed correlation matrices across all pairs of neurons for self and gaze tuning functions. Sorting the self matrices using hierarchical clustering (dendrogram with centroid linkage) and using the same indices to sort the entries of the gaze matrix revealed markedly different patterns across the two representations (**Figure 4G**). Consistent with visual intuition, the correlation between corresponding entries of the self and gaze matrices was statistically significant but small ($r^2$ = 0.02, p < 0.001; scatter plot in **Figure 4H**). A test for bimodality showed a unimodal distribution of agent preference index (p > 0.05, Hartigan's dip test; histogram in **Figure 4H**), suggesting that self and gaze representations arise from overlapping neuronal populations rather than distinct pools.

To determine how self and gaze representations are arranged in neural space, we applied the same subspace analysis used for self, prey and predator (**Methods**). If gaze and self maps occupy distinct neural axes, then the principal components derived from self maps should capture little variance of gaze maps. We found that the top ten self-PCs explained 65.7% of self variance but only 17.2% of gaze variance (**Figure 4I**). To determine whether this limited cross-variance reflected genuine shared structure or sampling noise, we performed a half-split cross-validation analysis of the covariance structure. For each iteration, we randomly divided the self maps into two halves to estimate a within-self noise floor, and computed matched covariance matrices for gaze maps. We then correlated the corresponding entries of the self and gaze covariance matrices across splits (500 repetitions). Across splits, self gaze covariance correlations were small but reliably above zero ($r^2$ = 0.02, p < 0.001), indicating the presence of a weak but consistent shared structure beyond noise. Thus, self and gaze representations are mostly distinct but not entirely orthogonal.

Therefore, we quantified the angles between the subspace representations, using the alignment index (computed as the variance of gaze tuning functions captured by the top ten self-PCs normalized by the variance captured by the top ten gaze-PCs). Values near zero indicate orthogonality, whereas values near one indicate collinearity among subspaces. The alignment index between self and gaze was greater than zero, but significantly below shuffled control (0.24, p < 0.001; **Figure 4J**), meaning that self and gaze share little variance but are more orthogonal than expected by chance. That is, their neural subspaces are semi-orthogonal (Johnston et al., 2024). Finally, to rule out the possibility that gaze location simply reflected prey location, we computed the alignment between gaze and prey maps. These values were similarly low (chosen prey-gaze = 0.19; unchosen prey-gaze = 0.18), confirming that gaze occupies a distinct subspace rather than acting as a confound of prey representations.

### Subspaces for gaze and positions are also distinct but linearly transformable

Given this organization, we asked whether self and gaze subspaces could be separated into mutually orthogonal dimensions. Solving the generalized eigenvalue problem identified distinct self- and gaze-specific dimensions (**Figure 4K**). Projecting maps across subspaces explained only a minimal portion of variance, confirming that each subspace captures variance specific to its corresponding variable (**Figure 4K**). Finally, we tested whether these orthogonal subspaces are nonetheless linearly related. Using projections onto the self subspace to predict gaze subspace activity via a linear transformation revealed significant decoding performance (**Figure 4L-M**). Cross-validated $R^2$ values exceeded shuffled controls, indicating that although self and gaze occupy separable neural subspaces, their geometry permits a systematic linear mapping between them. Thus, hippocampal representations of gaze and self position are organized in separable yet geometrically coordinated subspaces (cf. Elsayed et al., 2016; Yoo and Hayden, 2020).

This geometric organization corresponds to what Elsayed and colleagues (Elsayed et al., 2016) define as *orthogonal but linked* structure, in which separable subspaces share a consistent relational geometry.

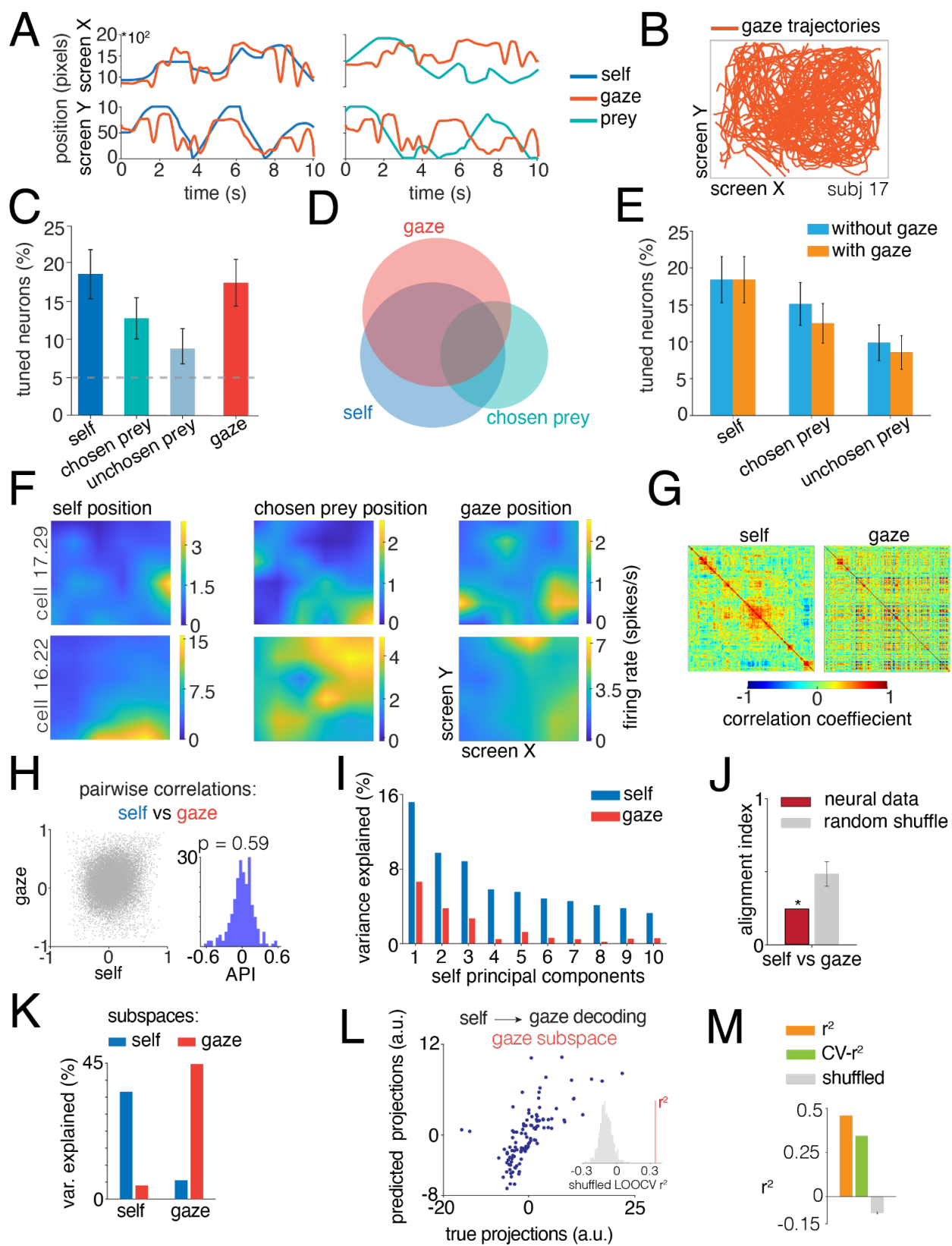

**Figure 4. Neural effects of gaze on hippocampal neurons. A.** Two illustrative examples of 10-second gaze trajectories (red line) aligned to the position of the self (blue line) and chosen

prey (cyan line). X- and Y-axes are separated. **B.** Gaze trajectory across a session for a typical patient. **C.** Using the LN-GLM, we find clear separable effects of gaze on neuronal firing in hippocampus, alongside main effects of the position of self and prey (compare with Figure 2A). **D.** Gaze effects did not affect a discrete population of neurons; instead, the populations largely overlapped. **E.** Including gaze as a regressor only modestly decreased the performance of models for coding of self and prey positions, indicating that the correlation between gaze and avatar position is not a major confounder. **F.** Example heatmaps show responses of two typical neurons to gaze position. For these two neurons, the self and chosen prey maps are also shown (compare with Figure 2E). **G.** Correlation matrices for self and gaze tuning functions across neurons. X and Y axes represent individual neurons, while the Z axis represents the strength of correlation between pairs of neurons. The order of the neurons is the same for the self and gaze matrix (dendrogram with centroid linkage applied to the self matrix only). **H.** Relationship between pairwise correlation values across self and gaze matrices (on the left). Each point represents a neuron pair. The weak correlation indicates that the similarity structure among neurons for self does not predict similarity structure among neurons for gaze. The histogram (on the right) shows the distribution of the epoch preference index, quantifying the strength of neural activity for self compared to that for gaze. The distribution is not significantly bimodal. **I.** Percent of self and gaze variance explained by the top ten self-PCs. **J.** Alignment index for neural and random data. Self and gaze representations are more orthogonal than expected by chance. **K.** Percent of variance explained by the self and gaze subspaces. Cross-maps projections significantly reduce the variance explained. **L.** Gaze subspace projections predicted from self subspace coordinates using a linear transformation, plotted against the observed gaze projections. Each point represents a spatial condition (36 spatial bins across the 4-dimensional subspace). Inset: shuffled cross-validated $R^2$ distribution; significance demonstrates that self and gaze subspaces are related by a linear transformation. M. Linear decoder performance: training $R^2$, leave-one-out cross-validation R2, and shuffled control for self-gaze linear mapping. Error bars denote 95% confidence interval of the shuffled distribution.

**Neural population geometry enables generalization across agents and gaze positions**

Our subspace analyses show that neural population activity for self, chosen prey, unchosen prey, and gaze occupies partially overlapping yet linearly transformable subspaces. This geometry suggests that the population code *could* support *abstraction* across agents, but the subspace analyses alone do not demonstrate it. Indeed, linear transformability does not imply that a decoder trained in one subspace will function in another, because the decoding axes may not be preserved by the transform. To test whether the geometry supports abstraction, we used cross-condition generalization performance (CCGP, Bernardi et al., 2020; **Figure 5A**).

We trained a linear classifier (SVM, **Methods**) on population activity for self position to discriminate left versus right screen locations, and then applied the same classifier to chosen prey and unchosen prey population activity to test for generalization across agents. When trained on self activity, the classifier generalized well to chosen prey (CCGP = 0.73; **Figure 5B**) and unchosen prey activity (CCGP = 0.32; **Figure 5B**), with both values far from the null distributions obtained by shuffling their respective class labels (two-tailed permutation test, both p < 0.001; **Figure 5B**). The value lower than 0.5 for unchosen prey is particularly interesting. It reflects an inverted decoding axis. In other words, while they are generalizable, the maps used for self-position and for unchosen prey are, generally speaking, reversed, much like photographic negatives. We observed a similar axis inversion when training the SVM on chosen prey related firing and tested on unchosen prey (CCGP = 0.26; p < 0.001; **Figure 5B**). (Note that we did not have enough data to test the predator condition).

To test whether gaze position follows the same generalization principle observed across agents, we trained a classifier on population activity associated with gaze location and evaluated its performance on self and prey related activity. The gaze activity-trained classifier generalized above chance to self (CCGP = 0.79; p < 0.001; **Figure 5C**) and chosen prey (CCGP = 0.62; p < 0.001; **Figure 5C**). Generalization to unchosen prey representations was not significant (CCGP = 0.44; p = 0.98; **Figure 5C**).

We quantified the angles between the decoding axes for each pair of agents and gaze representations by computing the angle between corresponding SVM weight vectors (**Figure 5D** and **Methods**). To visualize these relationships, we computed the pairwise cosine distances across weight vectors and projected them into a two-dimensional space using multidimensional scaling (MDS; **Figure 5D**). In this projection, the distances between points are proportional to the angular distances between decoding axes in neural space. Together, these results show that self, prey and gaze share coding axes that support abstraction across agents representations, with gaze and unchosen prey forming the only markedly orthogonal pair.

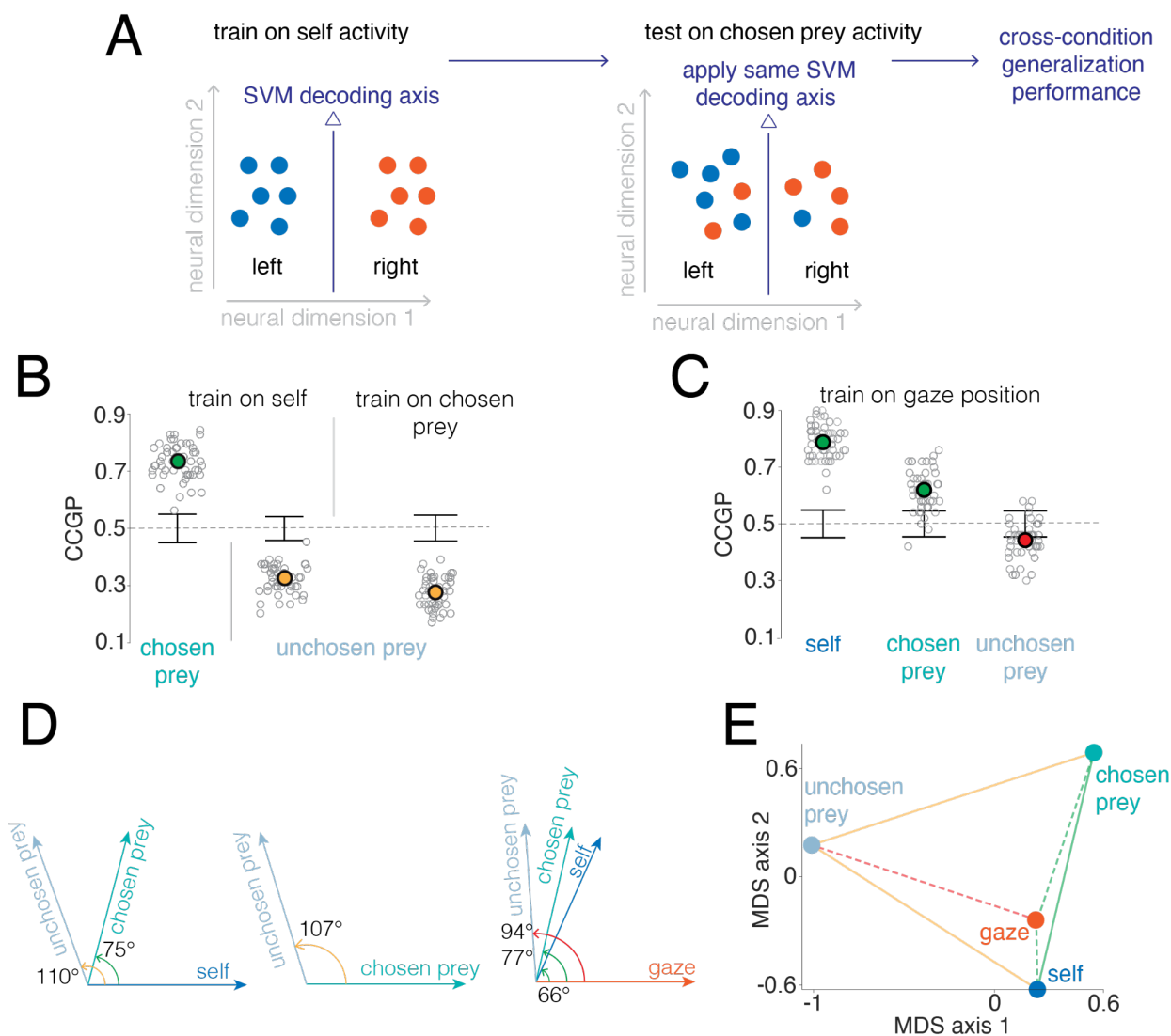

**Figure 5. Cross-condition generalization performance across agents and gaze. A.** Schematic of the CCGP analysis. A linear SVM is trained to classify left vs. right position on the screen using population activity for one agent (here, self). The same decoding axis is then applied to activity from a different agent (here, chosen prey), and CCGP quantifies how well decoding performance transfers across conditions. Here, "population activity" refers to the set of population firing rate vectors, each vector corresponding to the firing rates of all recorded neurons for a single spatial bin (either left or right) within a 40 seconds chunk of data (**Methods**). **B.** A linear SVM trained on self position activity generalized to chosen prey position and with inverted axis to unchosen prey position (left). A linear SVM trained on chosen prey position activity generalized with an inverted axis to unchosen prey position (right). Error bars represent ±2 standard deviations around chance accuracy (0.5), estimated from a null distribution generated by shuffling class labels. Gray circles represent CCGP values from individual resampling iterations (different random train/test splits), while colored circles represent their mean. Green indicates significantly above chance CCGP; yellow indicates significantly below chance CCGP (inverted axis generalization); red indicates no significant

CCGP. **C.** Same as in B, but here the classifier was trained on gaze bin wise population firing rate vectors and tested on the self and prey population vectors. **D.** Angles between decoding axes of linear classifiers each trained and tested separately on self, both prey and gaze, quantifying similarity of population coding directions. Decoding axes are determined as the SVMs weight vectors. **E.** Representation of the neural geometry obtained by applying multidimensional scaling to the set of pairwise cosine distances between decoding axes. The distance between points reflects the angular separation between decoding directions in the neural space. Links between points reflect the CCGP results. Continuous lines refer to CCGP across agents, while dashed lines refer to CCGP across agents and gaze. Color scheme as in panel B.

## DISCUSSION

We find, in a virtual prey pursuit task, that neurons in the human hippocampus can track positions of multiple moving items (self, two prey, and a predator) and of gaze, simultaneously. This multiplexing of information raises an *individuation problem*: if firing rate of a neuron changes, which of the items does that change refer to? Our data suggest that the hippocampus solves this problem by utilizing distinct transformable subspaces, rather than separate sets of neurons. This solution has several advantages. First, it enables the same neurons to participate in multiple representations, increasing coding capacity without requiring larger populations (Fusi et al., 2016; Rigotti et al., 2013; Tye et al., 2024). Second, subspace structure permits downstream circuits to flexibly read out task-relevant variables using simple linear operations (Cunningham & Yu, 2014; Elsayed et al., 2016; Mante et al., 2013). Third, organizing information into geometrically related subspaces preserves shared structure across agents, supporting abstraction (Johnston & Fusi, 2023), transfer, and perspective shifts, properties often attributed to hippocampal relational coding (Courellis et al., 2024; Eichenbaum, 2017). One implication of these abstract geometries is that spatial codes can generalize across agents. Thus, hippocampus may encode the relational structure of space in a way that can be reused across different entities in the environment: a computation learned for one agent, for instance estimating distance, can be applied to others, creating a flexible code for behavior (Behrens et al., 2018; Whittington et al., 2020; Rigotti et al., 2013). Together, these findings suggest that principles of manifold statistics apply to navigational codes in hippocampus just as they do to prefrontal representations (Fenton et al., 2010; Fenton, 2024; Levy et al. 2023; Nakai et al., 2024; Tang et al., 2023; Zhang et al., 2023).

Our observation that gaze position is encoded at the single-neuron level with a prevalence comparable to place coding suggests a new resolution to an important debate about the primate hippocampus (Hoffman et al., 2013; Mao et al., 2021; Martinez-Trujillo, 2025; Nau et al., 2018; Payne & Aronov, 2025; Piza et al., 2024; Watrous & Ekstrom, 2014; Wirth et al., 2017; Yoo et al., 2020). Prior work in nonhuman primates and humans has demonstrated hippocampal sensitivity to gaze and view, often interpreted as reflecting "spatial view" or scene-based coding that complements traditional place fields (Killian et al., 2012; Nau et al., 2018; Rolls et al., 1999). Conversely, others have argued that human (and more generally, primate) hippocampus may track position, much like that of rodents (Ekstrom et al., 2003; Kunz et al., 2021; Stangl et al., 2021). Our results suggest that the two views can be reconciled: both are robustly encoded - indeed, often in the same neurons - and using separate subspaces to differentiate them. This solution has the convenient benefit that it allows for ready transformation between gaze and spatial reference frame, which can facilitate embodied navigational decisions.

There are major ongoing debates about the nature of place cells in primates, and even questions about whether they exist (Mao et al., 2021; Martinez-Trujillo, 2025; Roberston et al., 1998; Rueckemann & Buffalo, 2017). While valuable, our results do not directly resolve these debates. First, our mapping functions are unusual in that players had a third-person view, rather than a first-person view, of the playing field (but, for relevance to place cells, see Mackay et al., 2024). Second, all movement was virtual, so vestibular and optic flow signals relevant to navigation were wholly lacking. Third, perhaps most critically, participants' gaze was not fixed, and, indeed, they typically fixated the self-avatar or the prey they pursued, so there was a close association between agents' position and gaze direction.

     A growing body of work argues that neural computations are best understood not at the level of single neurons but as trajectories evolving within low-dimensional manifolds embedded in high-dimensional population activity, with subspaces corresponding to task variables or latent dynamical modes (e.g., Gallego et al., 2017; Chung & Abbott, 2021; Perich et al., 2025). In motor and cognitive systems, such population-level structure has proven powerful for explaining flexibility, generalization, and learning, because behavior can be generated through the coordinated activation of a small set of neural modes rather than dedicated feature-specific neurons. Although the hippocampus has long been framed in representational, rather than dynamical, terms (e.g., place fields), manifold-oriented analyses suggest that its activity likewise occupies structured low-dimensional spaces, with geometric organization observable even in place-cell populations and other large-scale recordings (Fenton et al., 2010; Fenton, 2024; Nakai et al., 2024; Tang et al., 2023). Taking the manifold perspective for spatial encoding could be especially fruitful because spatial cognition inherently involves continuous variables, coordinate transforms, and multi-agent reference frames with computational motifs that are naturally expressed as aligned or partially overlapping subspaces within a shared neural geometry. Indeed, applying this approach to hippocampus for non-navigational codes has already yielded important insights (Bernardi et al., 2020; Boyle et al., 2024; Courellis et al., 2024; Esparza et al., 2023; Feather & Chung, 2023; Nieh et al., 2021; Whittington et al., 2020). As large-scale recordings become routine, adopting population-geometric frameworks may therefore provide a principled way to unify classic spatial coding phenomena with modern theories of abstraction, generalization, and flexible cognitive maps (Chung and Abbott, 2021).

**METHODS**

**Human intracranial neurophysiology**

Experimental data were recorded from 21 adult patients (6 males and 15 females) undergoing intracranial monitoring for epilepsy. The hippocampus was not a seizure focus area of any patients included in the study. Single neuron data were recorded from stereotactic (sEEG) probes, specifically AdTech Medical probes in a Behnke-Fried configuration. Each patient had an average of 3 probes terminating in left and right hippocampus. Electrode locations are verified by co-registered pre-operative MRI and post-operative CT scans. Each probe includes 8 microwires, each with 8 contacts, specifically designed for recording single-neuron activity. Single neuron data were recorded using a 512-channel Blackrock Microsystems Neuroport system sampled at 30 kHz. To identify single neuron action potentials, the raw traces were spike-sorted using the WaveClus sorting algorithm (Chaure et al., 2018) and then manually evaluated. Noise was removed and each signal was classified as multi or single unit using several criteria: consistent spike waveforms, waveform shape (slope, amplitude, trough-to-peak), and exponentially decaying ISI histogram with no ISI shorter than the refractory period (1 ms). The analyses here used only single unit activity.

**Electrode visualization**

Electrodes were localized using the software pipeline intracranial Electrode Visualization (iELVis, Groppe et al., 2017) and plotted across patients on an average brain using Reproducible Analysis & Visualization of iEEG (RAVE, Magnotti et al., 2020). For each patient, DICOM images of the preoperative T1 anatomical MRI and the postoperative Stealth CT scans were acquired and converted to NIfTI format (Li et al., 2016). The CT was aligned to MRI space using FSL (Jenkinson and Smith, 2001; Jenkinson et al, 2002). The resulting coregistered CT was loaded into BioImage Suite (version 3.5β1) (Joshi et al., 2011) and the electrode contacts were manually localized. Electrodes coordinates were converted to patient native space using iELVis MATLAB functions (Joshi et al, 2011; Yang et al., 2012) and plotted on the Freesurfer (version 7.4.1, Dale et al., 1999) reconstructed brain surface. Microelectrode coordinates are taken from the first (deepest) macro contact on the Ad-Tech Behnke Fried depth electrodes. RAVE was used to transform each patient's brain and electrode coordinates into MNI152 average space (Magnotti et al., 2020). The average coordinates were plotted together on a glass brain with the hippocampus segmentation and colored by patient.

**The prey-pursuit task**

**Task description.** The task used here is similar to one we have previously used in macaques (Yoo et al., 2020 and 2021) and the same as the one we have used in humans in Chericoni et al., 2025. At trial start, two or three shapes appeared on a gray computer monitor placed directly in front of the subject. The yellow circle (15-pixels in diameter) was an avatar that represented the subject and began at the center of the screen. Subject position was determined by the joystick and was limited by the screen boundaries. A square (30 pixels in length) represented the prey(s). Prey movement was determined by a simple algorithm (see below). Each trial ended with either the successful capture of the prey or after 20 seconds, whichever came first. Successful capture was defined as any spatial tangency between the avatar circle and the prey square. Capture resulted in an immediate numeric reward. Prey movement was generated interactively using an A-star algorithm (Hart et al., 1968). Specifically, for every frame (16.67ms), we computed the cost of 15 possible future positions the prey could move to in the next time-step. These 15 positions were equally spaced on the circumference of a circle centered on the current position of the prey, with a radius equal to the maximum distance the prey could travel within one time-step. The cost was based on two factors: the position in the field and the position of the subject's avatar. The field that the prey moved in had a built-in bias for cost, which made the prey more likely to move toward the center. The cost due to distance from the avatar of the subject was transformed using a sigmoidal function: the cost became zero beyond a certain distance so that the prey did not move, and it became greater as distance from the avatar of the subject decreased. From these 15 positional costs, the position with the lowest cost was selected for the next movement. If the next movement was beyond the screen range (1,800 × 1,000 resolution), then the position with the next lowest cost was selected until it was within the screen range. The maximum subject speed was 23 pixels per frame (each frame = 16.67ms).

To ensure sufficient time of pursuit, the minimum distance between the initial position of each subject avatar and prey was 400 pixels. The maximum and minimum speeds of the prey remained the same across subjects. In trials with predators, a predator (triangle shape) appeared on 50% of trials. Capture by the predator led to points loss. Predators came in five different types (indicated by color) indicating different levels of points loss, ranging from 1 to 5 points. The algorithm of the predator is to minimize the distance between itself and player. Unlike the prey, the predator algorithm is governed by this single rule. The design of the task reflects primarily the desire to have a rich and variegated virtual world with opportunities for choices at multiple levels that is neither trivially simple nor overly complex.

**Experimental apparatus.** Patients played at least 100 trials (average 119 trials) of the prey-pursuit task using a joystick (**Figure 1A, B**). The joystick was a commercially available joystick with a built-in potentiometer (Logitech Extreme Pro 3D). The joystick position was read out by a custom-coded program in Matlab running on the stimulus-control computer. The joystick was controlled by an algorithm that detected the positional change of the joystick and limited the maximum pixel movement to within 23 pixels in 16.67ms. Task events were synchronized to the neural recording system via comments, sent through analog port, from the computer playing the task to the Neural Signal Processor (NSP) at 30 kHz.

**Eye Tracking**

A subset of six patients performed the task while wearing a binocular, head-mounted eye-tracking system (ISCAN, Inc.) that recorded the patient's field of view (scene camera) at 60 Hz and eye movements via two eye cameras. Before the experiment, we performed a point-of-regard calibration procedure to map the patients' eye positions to the matrix of the head-mounted scene camera (640 × 480 pixel space). The calibration was performed on the same monitor that the patients viewed during the experiment. Patients were instructed to view five calibration points on the computer monitor, consisting of a center point and four outer points. The eye-calibration software synchronized the monocular eye-movement data with the scene camera, superimposing a crosshair indicating the center of gaze onto the scene video frames in real-time. If the participant looked outside the field of view of the scene camera frames, gaze position could not be estimated, and no crosshair was displayed on the corresponding frame.

We recorded scene camera videos using a CORENTSC DVR (I.O. Industries). This DVR recorded all videos at 60 frames per second and sent pulses to the Blackrock Cerebus neural recording system (neural signal processor; NSP) for every captured frame, as well as the start and end of video recording, thus synchronizing video data to neural activity by providing a timestamp (in neural time) for every frame. Eye position data was synchronized to neural activity via serial cable going directly from the eye tracking computer to the NSP.

To reconstruct gaze trajectories within the virtual environment, we utilized DeepLabCut (Mathis et al., 2018) to train a neural network to automatically label relevant objects in the frames. These included the crosshair (patient's point of gaze) as well as eight landmarks defining the monitor geometry: the four screen corners and the midpoints of the top, bottom, left, and right edges. The DeepLabCut output included the location coordinates for these designated landmarks and crosshair found in each frame. Using these landmarks, a projective geometric transformation (homography) was computed for each video frame via MATLAB's *fitgeotform2d* function. The resulting transformation matrix was applied to the crosshair coordinates to map the gaze position from the head-mounted camera's frame of reference to the task's monitor coordinate system, allowing us to determine gaze position throughout the experiment and identify moments when the gaze was following either prey (**Figure 4A**, **B**).

**Linear-nonlinear model (LN-GLM)**

To test the selectivity of neurons for spatial experimental variables, we constructed GLMs with navigational variables (Hardcastle et al., 2017; Pillow et al., 2008). The GLMs estimated the spike rate of one neuron during time bin $t$ as an exponential function of the weighted sum of the relevant value of each variable at time $t$, for which the weights are determined by a set of coefficients $w_i$. The estimated firing rates from the GLMs can be expressed as follows:

$$r = e^{(\sum X_i^T w_i)/dt}$$

Here, $r$ denotes the predicted firing rate across $T$ time points for a single neuron, and the variables of interest were the 2D positions of the self, both prey, the predator, and gaze, each expressed as an XY coordinate on the screen. This formulation allowed us to model neural activity continuously without time-locking to specific events. Each position variable was discretized into a 6×6 grid (36 bins), and each grid was represented using one-hot encoded columns in the design matrix $X$. Thus, each neuron was assigned a design matrix whose rows corresponded to the number of samples (concatenated across successful trials) and whose columns corresponded to the total number of spatial bins across variables (e.g., predictors being self + prey position, then $X$ would have 36×2 columns).

Unlike conventional tuning-curve analysis, GLMs do not assume a parametric shape for tuning. Instead, the weights defining each 2D tuning surface were optimized by maximizing the Poisson log-likelihood of the observed spike train. Because all predictors were 2D grids, we applied separate smoothness regularization to the parameters of each variable, enforcing smoothness across both spatial dimensions and reducing overfitting.

Regularization strengths were selected by 10-fold cross-validation, and optimization used Matlab's fminunc with analytic gradients and Hessians. For each test fold, model performance was quantified by the spike-normalized increase in log-likelihood relative to a null model predicting only the mean firing rate of the training data. To characterize predictive accuracy, we also computed the correlation between predicted and empirical firing rates after Gaussian smoothing.

**Forward model selection.** Model selection followed a forward stepwise procedure based on cross-validated log-likelihood. We first fit all single-variable models, one for each spatial predictor. The best single-variable model was identified as the one producing the largest spike-normalized increase in log-likelihood relative to a null model that predicted only the mean firing rate. Next, we evaluated models containing two variables, but only those that included the best single variable. A more complex model was retained only if it produced a significant improvement in cross-validated log-likelihood relative to the simpler model (Wilcoxon signed-rank test across folds, one-tailed, α = 0.05). This procedure continued in the same manner for models with additional spatial variables: a model was accepted only if it significantly outperformed the best model from the previous step. If adding further variables failed to provide a significant improvement at any stage, the forward search terminated early. After the final model was identified, we also compared its log-likelihood increase against the null model baseline; neurons whose best model failed to exceed the null model significantly were classified as not tuned to any of the spatial variables considered.

**Spatial similarity index (SPAEF)**

To compare the similarity between two spatial representations, we used the spatial efficiency measure (SPAEF) that prior literature suggests being more robust than the 2D spatial correlation (Koch et al., 2018). It quantifies the similarity between two maps as follows:

$$SPAEF = 1 - \sqrt{(A-1)^2 + (B-1)^2 + (C-1)^2}$$

where $A$ is the Pearson correlation between two maps, $B$ is the ratio between the coefficients of variation for each map and $C$ is the activity similarity measured by histogram profiles. Values near −1 indicate anticorrelated maps (one tends to be high when the other is low), 0 indicates uncorrelated maps and 1 indicates perfect matching between the two. Although SPAEF is not mathematically constrained to [−1, 1], values outside this range are uncommon and did not occur in our dataset.

To quantify the similarity between self, prey and predator representation for each neuron map, we computed SPAEF separately for each neuron and each pair of agents (see "Data preprocessing" for spatial map construction). This produced one SPAEF value per neuron per agent pair. To determine if the SPAEF values were significantly different from zero, we built a shuffled distribution over 1000 permutations. We then used a two-sided test to extract p-values, considering significance at p<0.05. To determine the noise ceiling, we performed 1000 within-agent half-splits. That is, for each iteration and each neuron, we randomly split the session concatenated trials into two independent halves, computed the spatial tuning maps separately for each half and each spatial variable (self, prey and predator positions) and then measured the spatial similarity between the two resulting maps. This process estimates the upper bound of explainable variance by accounting for the inherent noise in the data, providing a reference against which observed spatial similarities can be compared.

**Neural subspaces analysis**

**Data preprocessing.** We computed a 2-dimensional spatial map for each neuron and each spatial variable (x and y coordinates of self, chosen prey, unchosen prey, predator and gaze position on the screen used to perform the task) and used these maps to construct population activity matrices. Specifically, the screen field was divided into a 6x6 grid (36 bins).

To estimate the spatial map of each neuron, we first concatenated all trials (ended with successful capture of the prey) into a continuous time series of the agent's x and y coordinates and the corresponding spike counts. For each bin, we extracted all time points whose (x, y) position fell within that bin and computed the mean firing rate across those samples. This produced a 6×6 spatial map for each neuron and each spatial variable. Maps were smoothed with a Gaussian filter (σ = 0.5) and vectorized into 36-element column vectors ("neuron spatial maps"). Stacking these vectors across neurons and participants, for each agent (that is, each spatial variable) we obtained a pseudo-population activity matrix $M_{agent} \in \mathbb{R}^{N \times 36}$ where $N$ is the number of recorded neurons. Because neurons differ widely in mean firing rate and response scale, each row (corresponding to one neuron) was z-scored across spatial bins, so that each row had zero mean and unit variance. We used these matrices as input for all the subsequent subspace analysis.

**Correlation matrices visualization.** To visualize the similarity structure across neurons, we computed neuron-neuron correlation matrices for each spatial variable (**Figure 3A**, **G**). Starting from the population activity matrix ($M_{agent}$), we extracted the Pearson correlation coefficient between all pairs of neurons, yielding an $N \times N$ correlation matrix ($C_{agent}$) in which each entry reflects the similarity of two neurons' spatial tuning profiles. To compare the correlation structure across agents (self, prey, predator) and gaze, we ordered neurons using a hierarchical clustering algorithm (dendrogram with centroid linkage applied to one reference matrix (for instance, self) and used the resulting neuron order to display all other agents' correlation matrices. This enabled visual comparison of the population correlation structure across agents and gaze. To quantify differences in correlation structure, we vectorized the unique pairwise neuron-neuron correlations (upper triangle without the diagonal) for each agent's matrix and computed the Pearson correlation between these vectors (**Figure 3B** and **Figure 4H**).

**Agent Preference Index (API).** To quantify whether single neurons activity was more strongly driven by one agent (or gaze) compared to another, we adapted the epoch-preference index of Elsayed and colleagues (Elsayed et al., 2016) to compute an *agent preference index*. For each neuron $i$, we first measured its tuning strength to each agent by taking the 95th-5th percentile range of its unnormalized spatial map for that agent ($S_{self}(i)$, $S_{chosen\,prey}(i)$, $S_{unchosen\,prey}(i)$, $S_{predator}(i)$, $S_{gaze}(i)$). Using unnormalized maps preserves absolute response magnitude, which is required for comparing tuning strengths. Because different agents (or gaze) can have different overall tuning magnitudes across the population, we normalized each neuron's tuning strength by the mean tuning strength across all neurons for that agent or gaze ($\overline{S}_{self}$, $\overline{S}_{chosen\,prey}$, $\overline{S}_{unchosen\,prey}$, $\overline{S}_{predator}$, $\overline{S}_{gaze}$). The API for neuron $i$ was then defined as the difference between these normalized tuning strengths. For instance, the API between self and prey for neuron $i$ would be:

$$API_{self-prey} = \frac{S_{self}(i)}{\overline{S}_{self}} - \frac{S_{prey}(i)}{\overline{S}_{prey}}$$

Positive values indicate stronger tuning for self relative to prey, and negative values indicate the opposite. Notably, a bimodal distribution of this index across neurons would indicate the presence of two neural subpopulations. For instance, one preferentially representing self position and another the prey position. We assessed bimodality using the Hartigan dip test (**Figure 3B** and **Figure 4H**).

**Principal components analysis.** To assess whether two or more agents and gaze representations shared a common low-dimensional structure, we performed principal component analysis (PCA). For ease of description, we describe this analysis using the self vs chosen prey comparison as an example, although the same procedure was applied to all other combinations of agents or gaze (**Figure 3C, D** and **Figure 4I**). We applied PCA to the zscored and transposed self population matrix (36 spatial bins x N neurons) and extracted the top ten self-PCs. We then projected the chosen prey population matrix onto these self-PCs and quantified the proportion of prey variance each self-PC captured. This revealed how much covariance is shared by self and prey representation.

**Cross-condition covariance correlation.** To determine whether the covariance structure of two representations reflected genuine shared structure or noise sampling, we compared cross-agent covariance against within-agent covariance, which provides an estimate of the noise ceiling. For ease of description, we refer to the self vs. chosen prey comparison, but the same procedure was applied to all other agent and gaze pairs. The within-agent baseline was generated as follows: for each participant, concatenated trials were randomly divided into two equal splits (500 iterations), and spatial maps for self were computed for each subset. These maps were concatenated across participants for each subset, yielding two independent estimates of the self representation. After z-scoring each subset, we computed the corresponding covariance matrices, and the correlation between these matrices provided the expected covariance similarity due purely to noise.

We then estimated cross-agent (self-prey) correlation using the same trial splits. For each iteration, covariance matrices for self and chosen prey were computed using the identical subset of trials, and their correlation was taken as that iteration's cross-agent similarity. The mean cross-agent correlation over 500 iterations was then compared to the within-self baseline. One-sided p-values were obtained as the fraction of iterations in which the within-self correlation exceeded the self-prey correlation, allowing us to assess whether the cross-agent similarity was greater than expected from sampling noise.

**Subspace overlap analysis.** To quantify how much of the neural population structure was shared between two agents, we computed the *alignment index* (Elsayed et al., 2016). For clarity, we describe the self vs. chosen prey comparison, but the same procedure was applied to all other agent and gaze pairs (**Figure 3E** and **Figure 4J**). We took the ten top self-PCs and asked how much of the chosen prey variance fell within this self-defined subspace. The index was computed as follow:

$$AI = \frac{tr(D_{self}^T C_{prey} D_{self})}{\sum_i^{10} \sigma_{prey}(i)}$$

Where $C_{prey}$ is the covariance of the prey population matrix ($M_{prey}$), $D_{self}$ contains the top ten self PCs and $\sigma_{prey}$ is the $i^{th}$ singular value of $C_{prey}$. The numerator quantifies how much prey variance lies in the self subspace; the denominator normalizes this by the maximum prey variance captured by a ten-dimensional subspace (that is, the prey variance captured by the top ten prey-PCs). Thus, the alignment index varies from 0 (subspace orthogonality) to 1 (subspace collinearity).

To determine whether the observed alignment was lower than expected given dimensionality and the overall covariance structure, we estimated the covariance of the combined self and prey activity and sampled 1,000 random ten-dimensional orthonormal bases from this covariance. Alignment indices were recomputed for each random subspace to form a shuffled baseline. One-sided p-values were defined as the fraction of random subspaces whose alignment index was smaller than the observed value, testing whether the neural subspaces were more orthogonal than expected by chance.

**Identifying orthogonal subspaces between agents.** To identify two mutually orthogonal subspaces that best capture self and chosen prey (or any other agents or gaze pair) variance, we adapted the orthogonal subspace optimization methods developed by Elsayed et al. (2016). Starting from the z-scored population activity matrices for self and prey ($M_{self}$ and $M_{prey}$) we computed the associated covariances ($C_{self}$ and $C_{prey}$). We then searched for two orthonormal bases $Q_{self} \in \mathbb{R}^{N \times d_{self}}$ and $Q_{prey} \in \mathbb{R}^{N \times d_{prey}}$ that maximized the sum of normalized variance captured in each subspace:

$$[\hat{Q}_{self}, \hat{Q}_{prey}] = argmax_{[Q_{self}, Q_{prey}]} \frac{1}{2} \left( \frac{tr(Q_{self}^T C_{self} Q_{self})}{\sum_{i=1}^{d_{self}} \sigma_{self}(i)} + \frac{tr(Q_{prey}^T C_{prey} Q_{prey})}{\sum_{i=1}^{d_{prey}} \sigma_{prey}(i)} \right)$$

Subject to $\quad Q_{self}^T Q_{prey} = 0, \quad Q_{self}^T Q_{self} = I, \quad Q_{prey}^T Q_{prey} = I$

$\sigma_{self}(i)$ and $\sigma_{prey}(i)$ denote the singular values of $C_{self}$ and $C_{prey}$. The dimensionalities $d$ for self and prey were chosen using an elbow criterion based on variance captured and decoding performance. The optimization was performed on the Stiefel manifold using a manifold-optimization toolbox (Boumal et al., 2014). This method simultaneously identifies the two subspaces, ensures complete orthogonality, and normalizes for differences in variance and dimensionality between agents (Cunningham and Ghahramani, 2014; Elsayed et al., 2016).

**Linear transformation between agents and gaze subspaces.** To test whether the inferred orthogonal subspaces for each agent were related by a consistent linear transformation, we projected the population activity onto each orthogonalized subspace. For each pair of agents and gaze (described here for self vs chosen prey), this projection yielded low-dimensional representation for self and prey ($X_{self} \in \mathbb{R}^{d_{self} \times C}$ and $X_{prey} \in \mathbb{R}^{d_{prey} \times C}$, with $C$ number of samples (spatial bins conditions). We then fit a linear decoder with least squares:

$$X_{prey} = W X_{self}$$

Where $W \in \mathbb{R}^{d_{prey} \times d_{self}}$ contains the regression weights mapping the activity in the self subspace onto the activity of the prey subspace. The quality of fit was quantified as:

$$R^2 = 1 - \frac{||X_{prey} - W X_{self}||_F^2}{||X_{prey}||_F^2}$$

$R^2$ were computed on the full data and using leave-one-out cross-validation across spatial bins. As a control, we generated a shuffled distribution by independently permuting the entries within each row of $X_{self}$ and refitting the decoder (1000 shuffles). The p-value was the fraction of shuffled CV $R^2$ values greater than or equal to the observed CV $R^2$ and tested whether the linear mapping between self and prey subspaces exceeded chance.

**Cross-correlation decoding performance (CCGP) analysis**

**Data preprocessing.** Because CCGP requires many independent population samples rather than a single trial averaged map per bin, we segmented the continuous neural recordings into multiple 40 seconds chunks to produce pseudo population firing rate vectors for each spatial bin and each chunk. For each patient, we concatenated spike trains and the x and y coordinates of each spatial variable across trials. We then divided the concatenated data into 40 seconds non overlapping chunks and computed firing rates for each of the 36 spatial bins within each chunk using a 6×6 grid. This resulted in a population activity matrix (neurons x 36 bins) for each chunk and each spatial variable.

**Defining classes for CCGP**. Because our goal was to test generalization across agents and gaze, we first had to define two categorical conditions that were comparable across self, both prey, and gaze. Since the x-coordinate is shared across spatial variables and can be discretized consistently across all spatial maps, we used left vs right spatial position on the screen as the decoding variables.

For each 40 seconds chunk derived population matrix, we identified the 24 (12 per side) spatial bins whose x-coordinates were the most leftward and the most rightward. For each side, among these 12 bins, we then selected the 6 bins with the highest occupancy. Notably, choosing 6 random bins from the left and right set yielded comparable CCGP, indicating that our results were not driven by a specific bin selection rule (data not shown). Each selected bin contributed one population activity vector (neurons × 1), and these vectors were stacked across bins, classes, and chunks. This produced a design matrix with a number of rows equal to 6 (bins) x 2 (left/right side) x number of chunks in a session, and a number of columns equal to the number of neurons ($D \in \mathbb{R}^{(n_{bins} \times n_{classes} \times n_{chunk}) \times n_{neurons}}$). This procedure yielded four matrices, one for self, one for chosen prey, one for unchosen prey, and one for gaze. Each matrix was paired with a binary class label vector indicating whether each population vector belonged to the left or right side of the screen. These matrices and label vectors were the inputs to the CCGP decoding analyses.

**CCGP computation.** For ease of description, we focus on the self vs. chosen prey generalization, though the same logic was applied to all agents and gaze pairs. The goal was to determine whether a left/right decoder learned from self-related activity could generalize to the population activity of the chosen prey. On each repetition (50 total), the self design matrix $D$ was split into a training set (70%) and a test set (30%), preserving the balance between left and right bins. A linear support vector machine (SVM; MATLAB's fitcsvm with 'KernelFunction', 'linear') was trained on the z-scored self training data; chosen prey test data were normalized using the mean and standard deviation computed from the self training set only. The trained decoder was then applied, without retraining, to the chosen prey test data (same left/right structure as the self test split). The CCGP for chosen prey was defined as the mean classification accuracy across repetitions.

**Shuffle baseline.** To determine whether the observed CCGP exceeded chance, we constructed a null distribution in which left/right labels were independently permuted across samples within each context (self, chosen prey, unchosen prey, gaze). For each shuffle (500 total), we reran the full 50 repetition procedure, yielding a distribution of shuffled CCGP values. The p-value was computed as the proportion of shuffled CCGP values that exceeded (or, for metrics below chance, were lower than) the observed CCGP.

**Decoding axes orientation and visualization in neuron space.** To compare how similarly different agents separated left/right position in neural space, we extracted one linear decoding axis per context (self, chosen prey, unchosen prey and gaze) and measured the angles between them. For each context, we started from the corresponding CCGP design matrix $D$ and the corresponding left/right labels, z-scored firing rates across samples (per neuron) and trained a linear SVM (MATLAB fitcsvm, 'KernelFunction','linear') to discriminate left/right. The resulting SVM weight vector (beta coefficients) defines a single decoding axis in neuron space for that context. We repeated this procedure separately for each agent and normalized each axis to unit length. We then quantified the similarity between any two decoders axes by computing the cosine of the angle between their weight vectors and converted these cosines to angles in degrees. These angles report the geometric separation between agents' left/right decision boundaries in the shared neuronal population space.

To visualize how self, chosen prey, unchosen prey and gaze encode left/right position in the shared neuronal population, we computed the cosine similarity between every pair of axes, yielding a 4×4 cosine matrix. We converted these cosines to pairwise distances using $D_{ij} = \sqrt{2(1 - \cos(\theta_{ij}))}$ and obtained a cosine distance matrix. We then applied multidimensional scaling (MDS) to this matrix and plotted the first two dimensions. The resulting 2D embedding provides a visualization of the relative geometry of the four decoding axes while preserving their pairwise angular relationships as closely as possible.